\documentclass[dvips]{article}
\usepackage{supertabular,lscape,epsfig}
\usepackage{amssymb}
\usepackage{amsmath}

\DeclareSymbolFont{ppa}{OT1}{ppl}{m}{it}
\DeclareMathSymbol{\vv}{\mathalpha}{ppa}{'166}

\begin{document}

\newcommand{\dd}{\,{\rm d}}
\newcommand{\ie}{{\it i.e.},\,}
\newcommand{\etal}{{\it et al.\ }}
\newcommand{\eg}{{\it e.g.},\,}
\newcommand{\cf}{{\it cf.\ }}
\newcommand{\vs}{{\it vs.\ }}
\newcommand{\zdot}{\makebox[0pt][l]{.}}
\newcommand{\up}[1]{\ifmmode^{\rm #1}\else$^{\rm #1}$\fi}
\newcommand{\dn}[1]{\ifmmode_{\rm #1}\else$_{\rm #1}$\fi}
\newcommand{\upd}{\up{d}}
\newcommand{\uph}{\up{h}}
\newcommand{\upm}{\up{m}}
\newcommand{\ups}{\up{s}}
\newcommand{\arcd}{\ifmmode^{\circ}\else$^{\circ}$\fi}
\newcommand{\arcm}{\ifmmode{'}\else$'$\fi}
\newcommand{\arcs}{\ifmmode{''}\else$''$\fi}
\newcommand{\MS}{{\rm M}\ifmmode_{\odot}\else$_{\odot}$\fi}
\newcommand{\RS}{{\rm R}\ifmmode_{\odot}\else$_{\odot}$\fi}
\newcommand{\LS}{{\rm L}\ifmmode_{\odot}\else$_{\odot}$\fi}

\newcommand{\Abstract}[2]{{\footnotesize\begin{center}ABSTRACT\end{center}
\vspace{1mm}\par#1\par
\noindent
{~}{\it #2}}}

\newcommand{\TabCap}[2]{\begin{center}\parbox[t]{#1}{\begin{center}
  \small {\spaceskip 2pt plus 1pt minus 1pt T a b l e}
  \refstepcounter{table}\thetable \\[2mm]
  \footnotesize #2 \end{center}}\end{center}}

\newcommand{\TableSep}[2]{\begin{table}[p]\vspace{#1}
\TabCap{#2}\end{table}}

\newcommand{\FigCap}[1]{\footnotesize\par\noindent Fig.\  %
  \refstepcounter{figure}\thefigure. #1\par}

\newcommand{\TableFont}{\footnotesize}
\newcommand{\TableFontIt}{\ttit}
\newcommand{\SetTableFont}[1]{\renewcommand{\TableFont}{#1}}

\newcommand{\MakeTable}[4]{\begin{table}[htb]\TabCap{#2}{#3}
  \begin{center} \TableFont \begin{tabular}{#1} #4 
  \end{tabular}\end{center}\end{table}}

\newcommand{\MakeTableSep}[4]{\begin{table}[p]\TabCap{#2}{#3}
  \begin{center} \TableFont \begin{tabular}{#1} #4 
  \end{tabular}\end{center}\end{table}}

\newenvironment{references}%
{
\footnotesize \frenchspacing
\renewcommand{\thesection}{}
\renewcommand{\in}{{\rm in }}
\renewcommand{\AA}{Astron.\ Astrophys.}
\newcommand{\AAS}{Astron.~Astrophys.~Suppl.~Ser.}
\newcommand{\ApJ}{Astrophys.\ J.}
\newcommand{\ApJS}{Astrophys.\ J.~Suppl.~Ser.}
\newcommand{\ApJL}{Astrophys.\ J.~Letters}
\newcommand{\AJ}{Astron.\ J.}
\newcommand{\IBVS}{IBVS}
\newcommand{\PASP}{P.A.S.P.}
\newcommand{\Acta}{Acta Astron.}
\newcommand{\MNRAS}{MNRAS}
\renewcommand{\and}{{\rm and }}
\section{{\rm REFERENCES}}
\sloppy \hyphenpenalty10000
\begin{list}{}{\leftmargin1cm\listparindent-1cm
\itemindent\listparindent\parsep0pt\itemsep0pt}}%
{\end{list}\vspace{2mm}}

\def\TYLDA{~}
\newlength{\DW}
\settowidth{\DW}{0}
\newcommand{\dw}{\hspace{\DW}}

\newcommand{\refitem}[5]{\item[]{#1} #2%
\def\REFARG{#3}\ifx\REFARG\TYLDA\else, {\it#3}\fi
\def\REFARG{#4}\ifx\REFARG\TYLDA\else, {\bf#4}\fi
\def\REFARG{#5}\ifx\REFARG\TYLDA\else, {#5}\fi.}

\newcommand{\Section}[1]{\section{\hskip-6mm.\hskip3mm#1}}
\newcommand{\Subsection}[1]{\subsection{#1}}
\newcommand{\Acknow}[1]{\par\vspace{5mm}{\bf Acknowledgements.} #1}
\pagestyle{myheadings}

\newfont{\bb}{ptmbi8t at 12pt}
\newcommand{\xrule}{\rule{0pt}{2.5ex}}
\newcommand{\xxrule}{\rule[-1.8ex]{0pt}{4.5ex}}
\def\thefootnote{\fnsymbol{footnote}}
\begin{center}
{\Large\bf The Optical Gravitational Lensing Experiment.
\vskip1pt
Planetary and Low-Luminosity Object Transits
\vskip1pt
in the Fields of Galactic Disk. Results of the 
\vskip4pt
2003 OGLE Observing
Campaigns\footnote{Based on observations obtained with the 1.3~m Warsaw
telescope at the Las Campanas Observatory of the Carnegie Institution of
Washington.}}
\vskip.6cm
{\bf
A.~~U~d~a~l~s~k~i$^1$,~~M.\ K.~~S~z~y~m~a~{\'n}~s~k~i$^1$,
~~M.~~K~u~b~i~a~k$^1$,~~G.~~P~i~e~t~r~z~y~\'n~s~k~i$^{1,2}$,
~~I.~~S~o~s~z~y~\'n~s~k~i$^{1,2}$,~~K.~~\.Z~e~b~r~u~\'n$^1$,~~O.~~S~z~e~w~c~z~y~k$^1$,
~~and~~\L.~~W~y~r~z~y~k~o~w~s~k~i$^1$}
\vskip2mm
$^1$Warsaw University Observatory, Al.~Ujazdowskie~4, 00-478~Warszawa, Poland\\
e-mail: (udalski,msz,mk,pietrzyn,soszynsk,zebrun,szewczyk,wyrzykow)@astrouw.edu.pl\\
$^2$ Universidad de Concepci{\'o}n, Departamento de Fisica,
Casilla 160--C, Concepci{\'o}n, Chile
\end{center}

\Abstract{We present results of two observing campaigns conducted by the
OGLE-III survey in the 2003 observing season aiming at the detection of new
objects with planetary transiting companions. Six fields of
${35\arcm\times35\arcm}$ each located in the Galactic disk were monitored
with high frequency for several weeks in February--July 2003. Additional
observations of three of these fields were also collected in the 2004
season. Altogether about 800 and 1500 epochs were collected for the fields
of both campaigns, respectively.

The search for low depth transits was conducted on about 230~000 stars with
photometry better than 15~mmag. It was focused on detection of planetary
companions, thus clear non-planetary cases were not included in the final
list of selected objects. Altogether we discovered 40 stars with shallow
(${\leq0.05}$~mag) flat-bottomed transits. In each case several individual
transits were observed allowing determination of photometric
elements. Additionally, the lower limits on radii of the primary and
companion were calculated.

From the photometric point of view the new OGLE sample contains many very
good candidates for extrasolar transiting planets. However, only the future
spectroscopic follow-up observations of the OGLE sample -- determination of
the amplitude of radial velocity and exclusion of blending possibilities --
may allow to confirm their planetary status. In general, the transiting
objects may be extrasolar planets, brown dwarfs, M-type dwarfs or fake
transits caused by blending.

All photometric data of objects with transiting companions discovered
during the 2003 campaigns are available to the astronomical community from
the OGLE {\sc Internet} archive.}{}

\Section{Introduction} 

Detection of planetary transits in the spectroscopically discovered
extrasolar planetary system HD209458 (Henry \etal 1999, Charbonneau
\etal 1999) empirically proved that the photometric method of detection
of extrasolar planets, called the ``transit method'', can be a
potentially important tool in planetary searches. However, in spite of
many efforts no extrasolar planet was detected in this manner
during the next few years. The transit method offers many advantages
compared to the traditional spectroscopic method. When the photometric
observations are combined with precise spectroscopy all important
parameters of a planet like its size and mass can be derived providing
extremely important data to studying in detail the structure of planets
and evolution of extrasolar planetary systems.

In 2002 the Optical Gravitational Lensing Experiment (OGLE) survey
announ\-ced results of two photometric campaigns aimed at photometric
discovery of low depth transits caused by extrasolar planets or other sort
of ``low-luminosity'' objects (Udalski \etal 2002abc, 2003). About 140
stars such stars were found among about 150~000 stars in two lines-of-sight
in the Galactic disk -- toward the Galactic center and the constellation of
Carina. It was obvious from the very beginning that the significant part
of the discovered transits may be caused by stellar size companions, part
can be an effect of blending of eclipsing stars with bright unresolvable
neighbors while in the small number of objects the transits could be of
planetary origin. Unfortunately, the photometry, which allows determination
of size, cannot unambiguously distinguish planetary or brown dwarf from
late M-type dwarf stellar companions as all of them may have the radii of
the order of 0.1--0.2~\RS\ (1--2~$R_{\rm Jup}$).

The OGLE transiting objects were subject of many follow up studies in the
past two years. The most important follow up observations of the OGLE
objects were high resolution spectroscopic observations for radial velocity
changes. In the past two years the planetary status of four of the OGLE
transiting companions was confirmed by measuring small amplitude radial
velocity variation with the same period and in appropriate phase as
resulting from the OGLE photometric orbits: OGLE-TR-56 (Konacki \etal
2003a, Torres \etal 2004a, Bouchy \etal 2004b), OGLE-TR-113 (Bouchy \etal
2004a, Konacki \etal 2004), OGLE-TR-132 (Bouchy \etal 2004a) and
OGLE-TR-111 (Pont \etal 2004). Another OGLE object -- OGLE-TR-10 -- also
very likely hosts an extrasolar planet although additional radial velocity
observations are necessary to fully confirm its planetary status (Konacki
\etal 2003b, Bouchy \etal 2004b). Additionally, the high resolution
spectroscopy allowed to rule out possible blending scenarios making the
confirmations very reliable and sound. It is worth noting that except for
OGLE-TR-132, which was at the edge of detection, all the remaining OGLE
planets were on the OGLE short lists of the most likely planetary
candidates based on their photometric properties. The sixth known
extrasolar transiting planet, TrES-1, has recently been found by the small
telescope wide field TrES survey (Alonso \etal 2004).

Detection of the first transiting planets allowed to significantly increase
the sample of extrasolar planets with precisely known masses and radii,
thus providing crucial empirical data for testing the planetary models etc.
Surprisingly the first three OGLE planets (OGLE-TR-56, OGLE-TR-113 and
OGLE-TR-132) turned out to form a new class of ``very hot Jupiters'' --
giant planets orbiting their hosts stars with periods shorter than
2~days. Such systems have been unknown so far in the solar neighborhood. It
is clear that the detection of additional transiting planets is of high
importance for extrasolar planetary science.

In 2003 the OGLE survey conducted two additional observational campaigns
aimed at the detection of additional transiting exoplanets. Altogether six
new fields located in the Galactic disk in the constellations of Carina,
Centaurus and Musca were monitored regularly with high time resolution for
a few weeks each. The search for objects with transiting companions was
performed in similar manner as in the past campaigns. However, it was more
focused on planetary transits. Hence, many clear detections of low mass
stellar transiting companions were removed from the final sample of objects
with transiting companions.

Similarly to our previous transit samples (Udalski \etal 2002abc, 2003) the
photometric data of our new objects with transiting companions from the
2003 fields can be found in the OGLE {\sc Internet} archive. Thus the
follow-up spectroscopic observations and confirmation of their planetary or
non-planetary status can be made in short time scale. Details and pointers
to the OGLE {\sc Internet} archive can be found at the end of this paper.

\Section{Observational Data}
Observations presented in this paper were collected with the 1.3-m Warsaw
telescope at the Las Campanas Observatory, Chile (operated by the Carnegie
Institution of Washington), equipped with a wide field CCD mosaic
camera. The camera consists of eight ${2048\times4096}$ pixel SITe ST002A
detectors. The pixel size of each of the detectors is 15~$\mu$m giving the
0.26~arcsec/pixel scale at the focus of the Warsaw telescope. Full field of
view of the camera is about ${35\arcm\times35\arcm}$. The gain of each chip
is adjusted to be about 1.3~e$^-$/ADU with the readout noise of about 6 to
9~e$^-$, depending on the chip. More details on the instrumental setup can
be found in Udalski (2003).
 
Two observing campaigns were conducted in 2003. The first of them -- OGLE
campaign \#3 -- started on February 12, 2003 and lasted up to March 26,
2003. The photometric data were collected during 39 nights spanning 43
days. Three fields of the Galactic disk were observed continuously with the
time resolution of about 15 minutes. Acronyms and equatorial coordinates
of these fields are provided in Table~1.

The second campaign -- OGLE campaign \#4 -- started on March 25, 2003.
While the main observing material was collected to the middle of May 2003,
the fields were also observed from time to time until July, 25,
2003. Moreover, additional but less extensive photometric monitoring of the
campaign \#4 fields was conducted during the 2004 observing season -- from
March 14, 2004 up to July 30, 2004. As in the previous campaigns three
fields of the Galactic disk were continuously monitored with the time
resolution of about 15 minutes. Details on the location of these fields and
their acronyms can also be found in Table~1.

\MakeTable{lcc}{12.5cm}{Equatorial coordinates of the observed fields}
{
\hline
\noalign{\vskip2pt}
\multicolumn{1}{c}{Field} & RA (J2000) & DEC (J2000)\\
\hline
\noalign{\vskip3pt}
\multicolumn{3}{c}{Campaign \#3}\\
\noalign{\vskip3pt}
CAR106  &  11\uph03\upm00\ups & $-61\arcd50\arcm00\arcs$ \\
CEN106  &  11\uph32\upm30\ups & $-60\arcd50\arcm00\arcs$ \\
CEN107  &  11\uph54\upm00\ups & $-62\arcd00\arcm00\arcs$ \\
\hline
\noalign{\vskip2pt}
\hline
\noalign{\vskip3pt}
\multicolumn{3}{c}{Campaign \#4}\\
\noalign{\vskip3pt}
CEN108  &  13\uph33\upm00\ups & $-64\arcd15\arcm00\arcs$ \\
MUS100  &  13\uph15\upm00\ups & $-64\arcd51\arcm00\arcs$ \\
MUS101  &  13\uph25\upm00\ups & $-64\arcd58\arcm00\arcs$ \\
\hline}

All observations were made through the {\it I}-band filter. The exposure
time of each image was set to 180 seconds. Altogether about 820 images were
collected for each field during our OGLE campaign \#3 and 1050 images in
2003 plus 460 images in 2004 for the campaign \#4 fields. The median seeing
of images in both campaigns was about~1\zdot\arcs2.

\Section{Data Reductions}

All collected images were preprocessed (de-biasing and flat-fielding) in
real time with the standard OGLE-III data pipeline (Udalski 2003).

Photometric reductions of the 2003 season images were performed off-line
after the end of both campaigns. Additional 2004 season data were reduced
with the same software setup in real time at the telescope. Similarly to
our previous campaigns the OGLE-III photometric data pipeline based on the
difference image analysis (DIA) method was used -- see details in Udalski
(2003).

Because no standard stars were observed during 2003 campaigns precise
calibration of the OGLE photometry of transit fields to the standard system
could not be obtained. However, part of the CAR106 field overlaps with
CAR$\_$SC1 field observed and well calibrated during the OGLE-II phase of
the OGLE survey. Based on the mean magnitudes of several bright stars
observed in both -- OGLE-II and OGLE-III -- the mean shift of the magnitude
scale between the OGLE-III magnitudes and OGLE-II calibrated data was
derived. Other fields were calibrated based on magnitudes of obtained in
this manner ``local'' standards in the CAR106 field. The error of the
magnitude scale should not exceed 0.1--0.15~mag.

Astrometric solution for the observed fields was performed similarly to the
previous campaigns (Udalski \etal 2002ac), \ie by cross-identification of a
few thousand brightest stars in each chip image with the Digitized Sky
Survey images of the same part of the sky. Then the transformation between
OGLE-III pixel grid and equatorial coordinates of the DSS (GSC) astrometric
system was calculated.

\Section{Search for Transits}

The search for objects with transiting companions was performed in almost
identical way as in the past campaigns.

Before the transit search algorithm was applied to the collected data, two
steps were performed. First a preselection procedure was run. We limited
our search for transiting objects to stars with very precise photometry,
\ie those with the {\it rms} from the entire time series collected in the
2003 season smaller than 15~mmag. Similarly to the 2002 Carina campaign we
did not make any preselection based on colors of stars. About 100~000 and
130~000 stars passed our ``good photometry'' cut in the campaign \#3 and
\#4 fields, respectively.

Next the photometric data of all objects were corrected for small scale
systematic effects using the data pipeline developed by Kruszewski and
Semeniuk (2003). Udalski \etal (2003) showed that this algorithm
efficiently removes various small scale systematic errors present in the
original data and makes it possible to detect lower depth transits.

Finally, all stars were subject to the transit search algorithm -- the BLS
algorithm of Kov{\'a}cs, Zucker and Mazeh (2002). Similar parameters of
the BLS algorithm as in the previous campaigns were used. The search for
transits was limited to periods from the range of 1.05~days to 10~days.

The final list was prepared after careful visual inspection of all light
curves which passed the BLS algorithm. The experience from our previous
campaigns and spectroscopic follow up observations indicates that the size
of exoplanets hardly exceeds the size of Jupiter. None of many larger size
transiting companions discovered by OGLE turned out to be a planet.
Therefore we set a tighter limit on the depth of transits than in the
previous campaigns and removed all objects that revealed transits deeper
than 0.05~mag.

It was also realized from the very beginning that when only two transits
had been observed during the campaign the derived periods may be uncertain.
Usually, the shortest period allowed by the distribution of epochs was
selected in such cases, but longer periods could not be excluded.
Spectroscopic follow-up observations of Bouchy \etal (2004b) showed that
indeed in a few cases so selected periods were inconsistent with
spectroscopy. The problem practically disappears when three transits are
observed. Therefore we left on our list only those objects for which three
or more transits were covered during our campaigns (this constraint was
lifted in two exceptions, \ie in the case of two good candidates in
campaign \#3 where only two transits were observed. The span of data
collected for campaign \#3 fields was much shorter than for the second
campaign making the detection of many transits more difficult).

To minimize the number of objects contaminating our final sample we removed
from the original list of detected stars a large number of small amplitude
events caused by grazing eclipses of regular stars of similar size and
brightness that produce V-shaped eclipses and large number of objects
revealing small depth eclipses but simultaneously a small amplitude
sinusoidal variation caused by distortion of the primary -- a clear sign of
a relatively massive companion (Udalski \etal 2002a, Drake 2003, Sirko and
Paczy{\'n}ski 2003). Also evident blends of faint eclipsing systems with a
brighter star were removed. However, it should be stressed that in the case
of more noisy light curves of fainter stars it is not easy to distinguish
between grazing eclipses and very non-central transits. Therefore, some of
the stars on our list might still be double stars, what can be easily
verified by the future spectroscopy.
 
The final periods of our candidates were found after careful examination of
the eclipse light curve -- by minimizing dispersion during the eclipse
phases that are very sensitive to period changes. The formal accuracy
of periods depends on the number of individual transits observed and the
total duration of observations. It is of the order of $10^{-4}P$ for the
only six week long campaign \#3 and $10^{-5}P$ for the two season long
campaign \#4 (when additional transits were observed in the 2004 season
data).

\Section{Results of the 2003 Campaigns}
Seventeen stars with small transiting companions from the fields of 2003
campaign \#3 have met our criteria. In the fields of 2003 campaign \#4
twenty three objects have been found.

\begin{figure}[htb]
\includegraphics[bb=30 40 505 310, width=12cm]{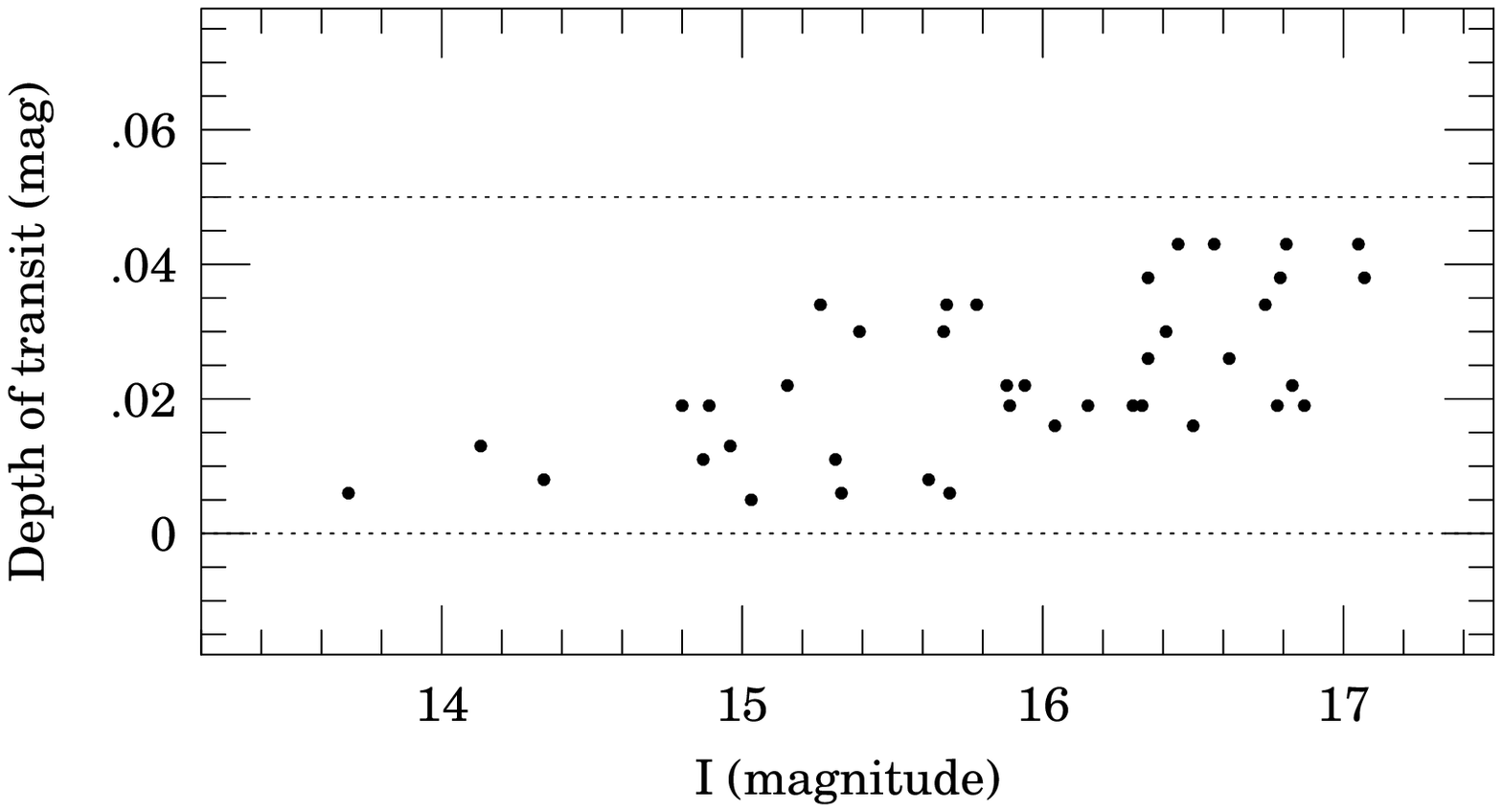}
\FigCap{Depth of transit \vs {\it I}-band magnitude of the host star.
Horizontal dotted line at the depth of 0.05 marks the limit of OGLE search.}
\end{figure}
\begin{figure}[htb]
\vglue-7mm
\includegraphics[bb=30 40 505 310, width=12cm]{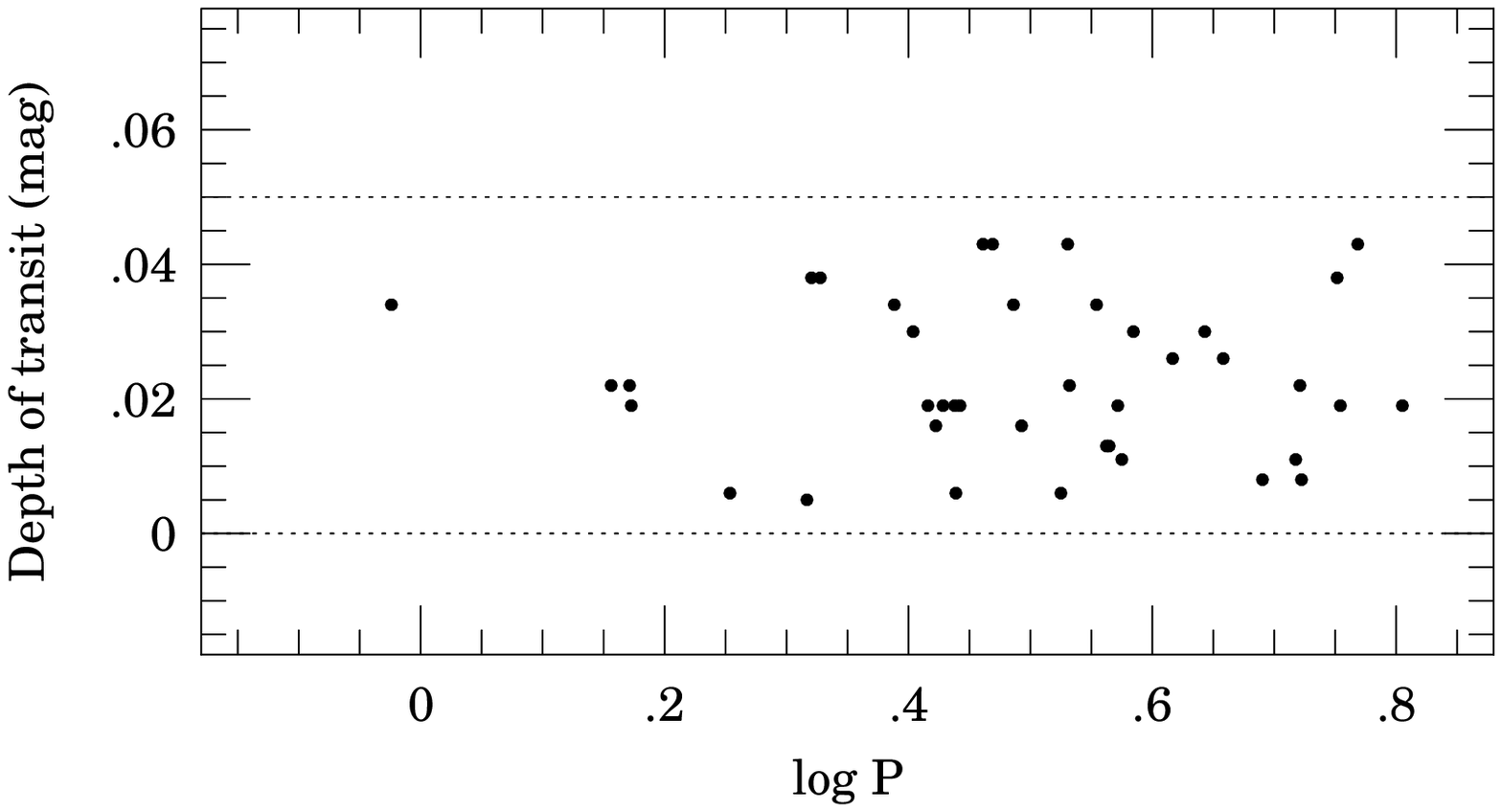}
\FigCap{Depth of transit plotted against the orbital period
($\log P$).}
\end{figure}
Similarly to our past campaigns we list the basic data on selected objects
in Table~2. We preserve our notation started in OGLE-III 2001 campaign: the
first object in Table~2 is designated as OGLE-TR-138. In Table~2 the
following data are tabulated: Identification, equatorial coordinates
(J2000), orbital period, epoch of mid-eclipse, {\it I}-band magnitude
outside transit, the depth of transit, number of transits observed ($N_{\rm
tr}$) and remarks. Accuracy of the magnitude scale is of about
0.1--0.15~mag. Although our search was limited to periods longer than
1.05~days, one object with shorter period (detected with the period of
$2P$) also entered the list.

In the Appendix we present the light curves and finding charts of selected
objects. For each object the full light curve and a close-up around the
transit are shown. It should be noted that the magnitude scale changes in
the close-up windows, depending on brightness, noise and transit depth. The
finding chart is a ${60\arcs\times60\arcs}$ subframe of the {\it I}-band
reference image centered on the star. The star is marked by a white
cross. North is up and East to the left in these images.

\MakeTableSep{l@{\hspace{6pt}}
c@{\hspace{5pt}}c@{\hspace{6pt}}c@{\hspace{5pt}}r@{\hspace{4pt}}
c@{\hspace{4pt}}c@{\hspace{4pt}}r@{\hspace{4pt}}l}{12.5cm}
{Objects with planetary or low luminosity transiting companions} 
{\hline
\noalign{\vskip4pt} 
\multicolumn{1}{c}{Name} & RA (J2000)  & DEC (J2000) &   $P$    &
\multicolumn{1}{c}{$T_0$}& $I$    &$\Delta I$  &  \multicolumn{1}{c}{$N_{\rm tr}$}&Rem.\\
           &             &             & [days]   & --2452000     &[mag] &[mag]  &\\ 
\noalign{\vskip4pt}
\hline
\noalign{\vskip4pt}
OGLE-TR-138 & 11\uph04\upm23\zdot\ups77 & $-61\arcd43\arcm10\zdot\arcs7$ & 2.64550 & 685.20281 & 16.04 & 0.016 & 3 & \\
OGLE-TR-139 & 11\uph02\upm43\zdot\ups79 & $-61\arcd37\arcm46\zdot\arcs7$ & 2.53420 & 684.64467 & 16.41 & 0.030 & 4 & \\
OGLE-TR-140 & 11\uph01\upm13\zdot\ups91 & $-61\arcd44\arcm21\zdot\arcs6$ & 3.39330 & 684.68281 & 16.45 & 0.043 & 4 & \\
OGLE-TR-141 & 11\uph01\upm52\zdot\ups09 & $-61\arcd58\arcm01\zdot\arcs4$ & 5.67860 & 683.75001 & 14.80 & 0.019 & 2 & \\
OGLE-TR-142 & 11\uph00\upm34\zdot\ups75 & $-62\arcd04\arcm09\zdot\arcs0$ & 3.06280 & 684.10926 & 15.68 & 0.034 & 4 & \\
OGLE-TR-143 & 11\uph34\upm03\zdot\ups83 & $-60\arcd54\arcm13\zdot\arcs0$ & 3.34980 & 682.92526 & 13.69 & 0.006 & 3 & \\
OGLE-TR-144 & 11\uph34\upm52\zdot\ups78 & $-60\arcd52\arcm35\zdot\arcs6$ & 2.44560 & 685.19335 & 16.74 & 0.034 & 5 & ST \\
OGLE-TR-145 & 11\uph33\upm04\zdot\ups34 & $-60\arcd34\arcm04\zdot\arcs2$ & 2.74120 & 685.48763 & 16.33 & 0.019 & 4 & ST \\
OGLE-TR-146 & 11\uph34\upm05\zdot\ups52 & $-60\arcd39\arcm38\zdot\arcs4$ & 2.94460 & 685.79072 & 16.57 & 0.043 & 5 & \\
OGLE-TR-147 & 11\uph32\upm30\zdot\ups45 & $-60\arcd38\arcm41\zdot\arcs6$ & 3.84170 & 686.11365 & 15.67 & 0.030 & 3 & \\
OGLE-TR-148 & 11\uph31\upm31\zdot\ups66 & $-60\arcd36\arcm52\zdot\arcs6$ & 1.43290 & 683.45961 & 16.83 & 0.022 & 7 & ST \\
OGLE-TR-149 & 11\uph30\upm45\zdot\ups10 & $-60\arcd43\arcm57\zdot\arcs9$ & 4.55170 & 684.68689 & 16.35 & 0.026 & 3 & \\
OGLE-TR-150 & 11\uph32\upm29\zdot\ups92 & $-60\arcd58\arcm25\zdot\arcs5$ & 2.07380 & 683.36717 & 15.03 & 0.005 & 6 & \\
OGLE-TR-151 & 11\uph32\upm21\zdot\ups84 & $-60\arcd50\arcm45\zdot\arcs5$ & 1.48350 & 683.46444 & 15.94 & 0.022 & 9 & \\
OGLE-TR-152 & 11\uph54\upm05\zdot\ups59 & $-62\arcd04\arcm38\zdot\arcs2$ & 3.73000 & 685.88809 & 16.30 & 0.019 & 4 & ST \\
OGLE-TR-153 & 11\uph55\upm21\zdot\ups60 & $-62\arcd03\arcm55\zdot\arcs3$ & 4.39560 & 685.15132 & 15.39 & 0.030 & 2 & \\
OGLE-TR-154 & 11\uph52\upm55\zdot\ups80 & $-62\arcd16\arcm53\zdot\arcs6$ & 3.66946 & 687.41198 & 14.96 & 0.013 & 4 & \\
\noalign{\vskip4pt}
\hline
\noalign{\vskip4pt}
OGLE-TR-155 & 13\uph33\upm26\zdot\ups00 & $-64\arcd16\arcm38\zdot\arcs4$ & 5.27700 & 701.09815 & 15.62 & 0.008 & 5 & \\
OGLE-TR-156 & 13\uph33\upm54\zdot\ups81 & $-64\arcd09\arcm43\zdot\arcs9$ & 3.58341 & 699.47628 & 15.26 & 0.034 & 8 & \\
OGLE-TR-157 & 13\uph34\upm22\zdot\ups90 & $-64\arcd07\arcm17\zdot\arcs5$ & 5.86821 & 698.99295 & 17.05 & 0.043 & 7 & \\
OGLE-TR-158 & 13\uph33\upm38\zdot\ups08 & $-64\arcd05\arcm23\zdot\arcs6$ & 6.38410 & 698.60150 & 15.89 & 0.019 & 5 & \\
OGLE-TR-159 & 13\uph31\upm54\zdot\ups49 & $-64\arcd02\arcm39\zdot\arcs6$ & 2.12676 & 697.54746 & 16.35 & 0.038 &16 & ST \\
OGLE-TR-160 & 13\uph30\upm42\zdot\ups95 & $-63\arcd57\arcm30\zdot\arcs6$ & 4.90185 & 698.32754 & 14.34 & 0.008 & 3 & \\
OGLE-TR-161 & 13\uph30\upm55\zdot\ups71 & $-63\arcd58\arcm38\zdot\arcs2$ & 2.74730 & 696.73100 & 15.69 & 0.006 & 4 & \\
OGLE-TR-162 & 13\uph31\upm01\zdot\ups09 & $-63\arcd58\arcm34\zdot\arcs9$ & 3.75819 & 731.37770 & 14.87 & 0.011 & 4 & \\
OGLE-TR-163 & 13\uph32\upm52\zdot\ups59 & $-63\arcd57\arcm44\zdot\arcs1$ & 0.94621 & 696.24111 & 15.78 & 0.034 &23 & \\
OGLE-TR-164 & 13\uph30\upm58\zdot\ups60 & $-64\arcd11\arcm53\zdot\arcs0$ & 2.68153 & 697.11292 & 16.15 & 0.019 &10 & \\
OGLE-TR-165 & 13\uph31\upm31\zdot\ups55 & $-64\arcd10\arcm51\zdot\arcs5$ & 2.89185 & 698.14260 & 16.81 & 0.043 & 9 & \\
OGLE-TR-166 & 13\uph32\upm23\zdot\ups86 & $-64\arcd10\arcm35\zdot\arcs2$ & 5.21920 & 699.83984 & 15.31 & 0.011 & 5 & \\
OGLE-TR-167 & 13\uph30\upm26\zdot\ups30 & $-64\arcd10\arcm09\zdot\arcs2$ & 5.26100 & 698.58614 & 15.88 & 0.022 & 4 & \\
OGLE-TR-168 & 13\uph32\upm12\zdot\ups07 & $-64\arcd32\arcm56\zdot\arcs2$ & 3.65080 & 696.94702 & 14.13 & 0.013 & 3 & \\
OGLE-TR-169 & 13\uph17\upm23\zdot\ups43 & $-64\arcd54\arcm54\zdot\arcs8$ & 2.76877 & 698.04287 & 16.78 & 0.019 &13 & \\
OGLE-TR-170 & 13\uph15\upm14\zdot\ups50 & $-64\arcd49\arcm28\zdot\arcs8$ & 4.13680 & 699.13261 & 16.62 & 0.026 & 7 & \\
OGLE-TR-171 & 13\uph13\upm52\zdot\ups00 & $-64\arcd41\arcm30\zdot\arcs0$ & 2.09180 & 696.49729 & 17.07 & 0.038 & 9 & \\
OGLE-TR-172 & 13\uph13\upm12\zdot\ups81 & $-64\arcd47\arcm13\zdot\arcs8$ & 1.79323 & 697.28151 & 15.33 & 0.006 & 8 & \\
OGLE-TR-173 & 13\uph14\upm56\zdot\ups11 & $-65\arcd02\arcm00\zdot\arcs4$ & 2.60590 & 698.87378 & 14.89 & 0.019 & 4 & \\
OGLE-TR-174 & 13\uph25\upm58\zdot\ups75 & $-65\arcd14\arcm50\zdot\arcs6$ & 3.11012 & 697.52261 & 16.50 & 0.016 & 7 & \\
OGLE-TR-175 & 13\uph27\upm47\zdot\ups31 & $-65\arcd14\arcm34\zdot\arcs8$ & 1.48830 & 697.01551 & 16.87 & 0.019 &11 & ST \\
OGLE-TR-176 & 13\uph25\upm33\zdot\ups73 & $-64\arcd55\arcm34\zdot\arcs1$ & 3.40478 & 700.66215 & 15.15 & 0.022 & 5 & \\
OGLE-TR-177 & 13\uph24\upm48\zdot\ups41 & $-64\arcd51\arcm37\zdot\arcs8$ & 5.64386 & 699.53686 & 16.79 & 0.038 & 3 & \\
\hline}

Fig.~1 presents the depth of transit plotted against {\it I}-band magnitude
of the host star. As expected the detection limit is roughly flat up to
$I\approx15.7$~mag and then rises slowly for noisier light curves of
fainter stars. In Fig.~2 the depth of transit is shown against $\log P$. No
relation or trends between these parameters are seen.

\Section{Discussion}

Forty new objects with small transiting companions were discovered in six
fields in the Galactic disk during two 2003 season OGLE campaigns.
Although the number of detected objects is smaller than in the past
campaigns our search in the 2003 season data was focused mainly on the
smallest companions which have larger probability to be extrasolar
planets. Therefore tighter selection criteria were applied and all
uncertain cases, objects with longer lasting transits -- evidently of
stellar origin, objects revealing significant variability between eclipses
suggesting more massive, stellar companions and other evident contaminators
were removed from the list of detected objects. Special care was taken to
detect transits as shallow as possible, as the probability of planetary
origin in such cases is larger. All existing detections of transiting
exoplanets indicate that the size of the extrasolar planets practically
does not exceed that of Jupiter so the depth of transits caused by planets
transiting FG-type stars must be of the order of 10~mmag or less.

Photometric observations of the transit shape allow to derive the size of
transiting companions. Unfortunately, the size alone cannot unambiguously
define the type of the transiting objects. Transits can be caused by
extrasolar planets or brown dwarfs or small late M-type dwarfs as all these
objects can have sizes of about ${R_{\rm Jup}}$. To distinguish between
these possibilities, radial velocity follow-up measurements are necessary
to estimate masses of companions. Another important contaminator of
photometrically selected transiting objects might be blending of a regular
totally eclipsing star with a close optical or physically related (wide
multiple system) unresolvable neighbor that can produce transit-like light
curve. Therefore some of our candidates listed in Table~2 can actually be
fake transits caused by blending effects.

Unfortunately, even in the clean unblended case additional information on
the size of the host star is necessary to derive the size of transiting
body, because the depth of transit provides only the ratio of those
sizes. Such additional information can come from moderate resolution
spectroscopy suitable for spectral classification, IR photometry (Dreizler
\etal 2002, Gallardo \etal 2004) or, in principle, from optical colors of
host stars. Unfortunately, in the case of the pencil beam survey of the
Galactic disk, like OGLE, significant and unknown interstellar extinction
makes dereddening of optical colors of individual stars very uncertain and
unreliable and thus of no use for estimation of stellar sizes.
 
When only single band photometry is available it is not possible to obtain
actual size of the companion when the errors of individual observations are
comparable to the transit depth, due to well known degeneracy between radii
of the host star and the companion, $R_s$, $R_c$, inclination, $i$, and
limb darkening, $u$. Photometric solutions of similar quality can be
obtained for different inclinations of the orbit and radii of components
(in the {\it I}-band the transit light curve is practically insensitive to
the limb darkening parameter $u$). Thus, the selection of the proper
solution is practically impossible. Only the lower limit of the size of the
companion can be calculated assuming that the transit is central, \ie
${i=90\arcd}$. The corresponding radius of the primary is in this case also
the lower limit.

In Table~3 we provide lower limits for the components radii calculated in
the same manner as in Udalski \etal (2002abc) assuming additionally that
the host star follows the mass-radius relation for main sequence stars
($R/\RS=(M/\MS)^{0.8}$). The resulting mass of the primary is also listed
in Table~3. In practice the transits might be non-central, thus the size of
the star and companion can be larger than given in Table~3. Also when the
host star is evolved the estimation may be inaccurate.

The estimations of the minimum sizes of transiting companions cannot be
considered as their actual dimensions as sometimes mistakenly
interpreted. The final parameters of the system can only be derived when
the spectroscopic observations are available. The aim of presenting
Table~3 is to provide information necessary for preselection of the most
promising planetary candidates from our 2003 transit sample. Objects with
the estimation of the minimum radius of transiting companion larger than
0.15--0.18~$R/\RS$ are almost certainly stellar binary systems and are
marked by ``ST'' in the remarks column of Table~2.

\renewcommand{\TableFont}{\footnotesize}
\MakeTable{lcccclccc}{12.5cm}{Lower limits of radii of stars and companions 
(central passage: $i=90\arcd$).}
{\cline{1-4}\cline{6-9}
\noalign{\vskip3pt}
Name       & $R_s$   & $R_c$ & $M_s$ &~~~~~& Name       & $R_s$   & $R_c$ & $M_s$\\
&[\RS]&[\RS]&[\MS]\\
\noalign{\vskip3pt}
\cline{1-4}\cline{6-9}
\noalign{\vskip3pt}
 OGLE-TR-138 &  0.74 &  0.082 &  0.69 &~~~~~&  OGLE-TR-158 &  0.29 &  0.035 &  0.21 \\
 OGLE-TR-139 &  1.00 &  0.150 &  1.00 &  & OGLE-TR-159 &  1.22 &  0.207 &  1.28 \\
 OGLE-TR-140 &  0.56 &  0.101 &  0.48 &  & OGLE-TR-160 &  1.02 &  0.081 &  1.02 \\
 OGLE-TR-141 &  0.79 &  0.095 &  0.75 &  & OGLE-TR-161 &  2.53 &  0.177 &  3.19 \\
 OGLE-TR-142 &  0.46 &  0.074 &  0.38 &  & OGLE-TR-162 &  1.02 &  0.092 &  1.03 \\
 OGLE-TR-143 &  0.49 &  0.034 &  0.41 &  & OGLE-TR-163 &  0.62 &  0.098 &  0.55 \\
 OGLE-TR-144 &  1.43 &  0.228 &  1.56 &  & OGLE-TR-164 &  1.12 &  0.134 &  1.15 \\
 OGLE-TR-145 &  1.82 &  0.219 &  2.12 &  & OGLE-TR-165 &  0.92 &  0.165 &  0.90 \\
 OGLE-TR-146 &  0.71 &  0.127 &  0.65 &  & OGLE-TR-166 &  1.31 &  0.118 &  1.40 \\
 OGLE-TR-147 &  1.05 &  0.158 &  1.06 &  & OGLE-TR-167 &  0.62 &  0.081 &  0.56 \\
 OGLE-TR-148 &  1.64 &  0.213 &  1.86 &  & OGLE-TR-168 &  0.77 &  0.077 &  0.73 \\
 OGLE-TR-149 &  0.91 &  0.128 &  0.89 &  & OGLE-TR-169 &  1.48 &  0.177 &  1.63 \\
 OGLE-TR-150 &  1.83 &  0.110 &  2.13 &  & OGLE-TR-170 &  1.23 &  0.172 &  1.29 \\
 OGLE-TR-151 &  1.02 &  0.132 &  1.02 &  & OGLE-TR-171 &  0.81 &  0.138 &  0.77 \\
 OGLE-TR-152 &  1.70 &  0.205 &  1.95 &  & OGLE-TR-172 &  1.19 &  0.083 &  1.24 \\
 OGLE-TR-153 &  0.79 &  0.119 &  0.75 &  & OGLE-TR-173 &  0.38 &  0.046 &  0.30 \\
 OGLE-TR-154 &  1.24 &  0.124 &  1.31 &  & OGLE-TR-174 &  0.75 &  0.083 &  0.70 \\
 OGLE-TR-155 &  1.42 &  0.113 &  1.54 &  & OGLE-TR-175 &  1.71 &  0.205 &  1.96 \\
 OGLE-TR-156 &  0.49 &  0.079 &  0.41 &  & OGLE-TR-176 &  1.02 &  0.132 &  1.02 \\
 OGLE-TR-157 &  0.88 &  0.158 &  0.85 &  & OGLE-TR-177 &  0.68 &  0.115 &  0.61 \\
\noalign{\vskip4pt}
\cline{1-4}\cline{6-9}}

Solid line in the close-up windows in the Appendix shows the transit model
light curve calculated for the central passage. In most cases the central
passage fit is practically indistinguishable from non-central so at this
stage it is impossible to derive other values than the lower limits of
radii provided in Table~3.
\vspace*{-1pt}

Because the search for transiting objects in the 2003 campaigns data was
focused on planetary systems the final list contains many objects with
companions for which the lower size limit is well within the planetary
range. Based on the shape of transits and photometric properties one can
select many very promising planetary transit candidates in our sample, for
instance OGLE-TR-158, OGLE-TR-162, OGLE-TR-164, OGLE-TR-167, OGLE-TR-172,
OGLE-TR-174, OGLE-TR-176. However, it cannot be excluded that other more
noisy or low transit depth systems also host a planet -- in such cases it
is difficult to assess the real transit shape.
\vspace*{-1pt}

Results of the past OGLE transit campaigns have shown how severe a
non-planetary background in the planetary transit searches can be. Very
often the OGLE sample of transit stars was misinterpreted as pure
``planetary'' candidates whereas the inclusion of many obvious stellar
companion objects was intentional as they could be very useful to derive
accurate masses and sizes of stars from poorly known lower part of the main
sequence (Bouchy \etal 2004b). Also it was clear that other contaminating
cases like blending or grazing eclipses can be present in the
sample. Spectroscopic follow up observations (Torres \etal 2004b, Torres
\etal 2005, Bouchy \etal 2004b) showed that this was indeed the case. In
particular, the blending turned out to be an important contaminating
factor, especially in the Galactic center line-of-sight where the eclipsing
stars from the Galactic bulge are very often blended with bright stars
mimicking low depth transits. This was one of the reasons to switch to the
much less crowded Galactic disk fields in the following OGLE campaigns.
\vspace*{-1pt}

The planetary transits have been detected so far with two different
approaches -- deep, narrow field surveys like OGLE and wide angle shallow
surveys like TrES. The former approach is more efficient as more stars can
be monitored simultaneously and allows to detect transiting planets
orbiting around faint stars of 13--17~mag located at much farther distances
from the Sun. On the other hand the wide field surveys can detect
transiting planets around brighter stars. This is often considered as an
advantage as more precise spectroscopy can be obtained, for example during
the transit, for analysis of the planetary atmosphere etc. Although it is
true now -- in the time scale of a decade next generation ${>25}$-m diameter
telescopes will diminish this gain. On the other hand, the estimations of
the number of transiting planets around bright stars available for wide
angle surveys indicate very limited number of such objects in the sky
(Pepper, Gould and DePoy 2003). Also, the accuracy of photometry must be
well below 10~mmag to allow detection of Jupiter-size planets transiting
FG-type stars (depth of transit of about 10~mmag) from the ground. Such an
accuracy is hard to achieve for wide angle small aperture
telescopes. Transits of HD209458 and TrES1 have relatively large depth
among all presently known transiting exoplanets.

Blending is the most important contaminating factor in both approaches.
While in the case of the deep narrow angle survey the most likely blending
scenarios are triple physically bound systems (except for the Galactic
bulge line-of-sight where by chance blends dominate), by chance blending in
low spatial resolution wide angle surveys is the main factor (due to small
angular resolution of detectors). Additional observational tests were
proposed to exclude potential blending contaminators or non-planetary cases
in the photometrically selected transiting candidates. However, the most
reliable tests can only be done with high resolution spectroscopy (Torres
\etal 2005). Other methods, for example multiband photometry, can only
provide weak, if any, constraints, in particular in the most important
cases when the transit is of small depth below 20~mmag.

Bouchy \etal (2004ab) showed that the high resolution spectroscopic follow
up observations of transit stars can be performed very efficiently provided
the large telescope with a high resolution spectrograph can work in the
multi-object mode being fed by fibers. In the cases of narrow angle surveys
like OGLE, where the selected objects are located in relatively small area
in the sky, spectra of many stars can be obtained simultaneously and follow
up of dozens of transit stars can be made in short time providing the
best observational material for full analysis of the transiting systems.

This approach seems to be the most effective and most complete. Therefore,
we did not attempt to clean further our sample listed in Table~2, believing
that this can be done in much more reliable way with multi-object
spectroscopic follow-up observations. While it is obvious that a part of
our sample will not be related to planetary companions we again remind that
the discovery of brown dwarfs among our candidates would shed a new light
on the so called ``brown dwarf desert'' problem (lack of brown dwarfs at
small orbits in binary systems) and allow for the first time to precisely
measure their masses and sizes. Also the mass-radius relation of the lower
part of the main sequence is poorly known, thus the determination of these
parameters in the case of any new low-mass stellar companion would be of
great value.

\Section{Data Availability}
The photometric data on the objects with transiting companions discovered
during the 2003 OGLE-III campaigns are available in the electronic form
from the OGLE archive:

\centerline{\it http://ogle.astrouw.edu.pl/} 

\centerline{\it ftp://ftp.astrouw.edu.pl/ogle/ogle3/transits/transits$\_$2003}

\noindent
or its US mirror

\centerline{\it http://bulge.princeton.edu/\~{}ogle}

\centerline{\it ftp://bulge.princeton.edu/ogle/ogle3/transits/transits$\_$2003}

\Acknow{The paper was partly supported by the Polish KBN grant
2P03D02124 to A.\ Udalski. Partial support to the OGLE project was provided
with the NSF grant AST-0204908 and NASA grant NAG5-12212 to
B.~Paczy\'nski. We acknowledge usage of the Digitized Sky Survey which was
produced at the Space Telescope Science Institute based on photographic
data obtained using the UK Schmidt Telescope, operated by the Royal
Observatory Edinburgh. A.~Udalski acknowledges support from the grant
``Subsydium Profesorskie'' of the Foundation for Polish Science.}

%\end{document}

\clearpage
\begin{figure}[p] 
\centerline{\large\bf Appendix}
\vskip23mm
\centerline{\includegraphics[width=13cm, bb=100 140 540 460]{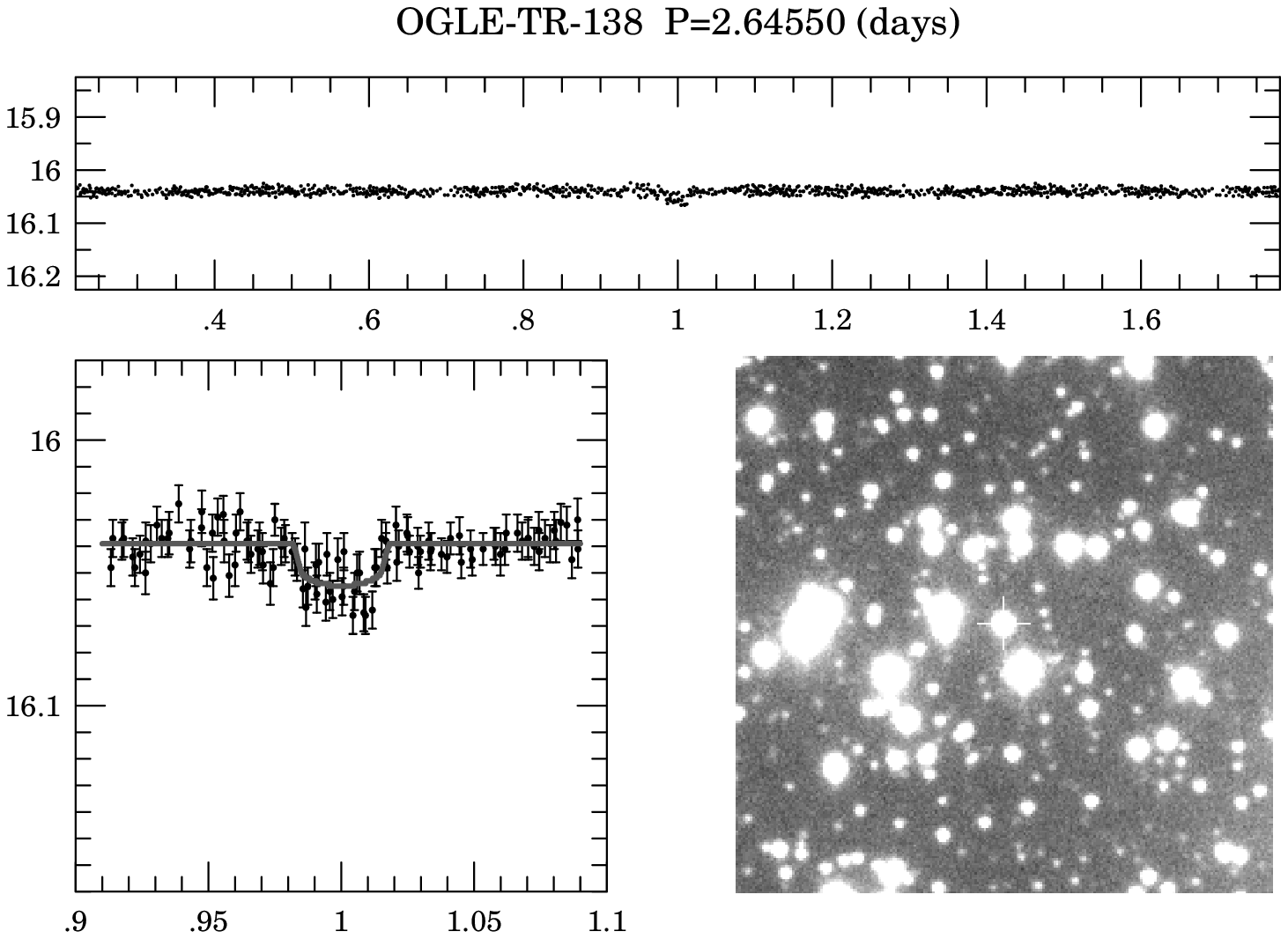}}
\vskip.4cm
\centerline{\includegraphics[width=13cm, bb=100 140 540 460]{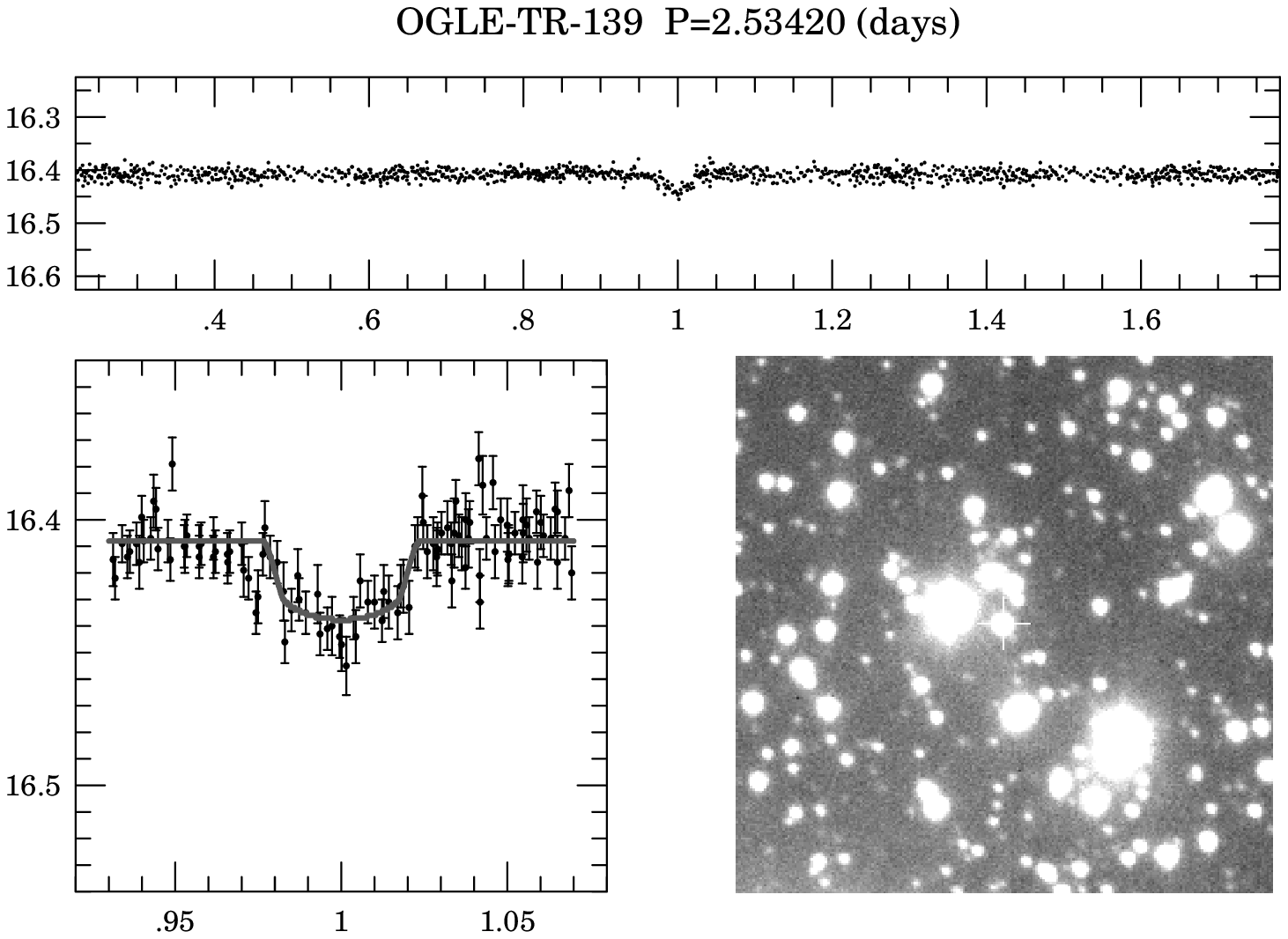}}
\end{figure} 
%\end{document}

\begin{figure}[p] 
\vglue17mm
\centerline{\includegraphics[width=13cm, bb=100 140 540 460]{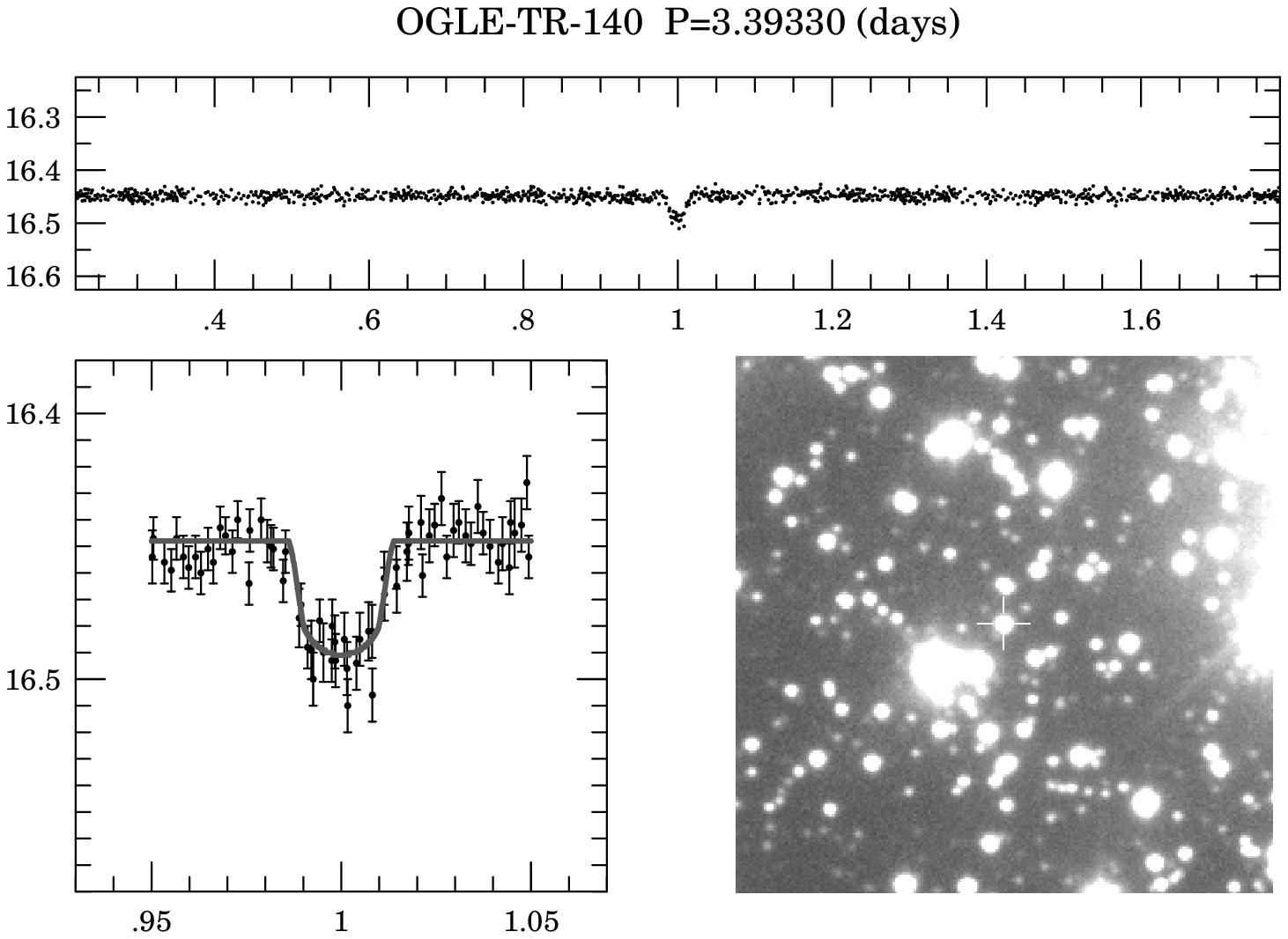}}
\vskip1.4cm
\centerline{\includegraphics[width=13cm, bb=100 140 540 460]{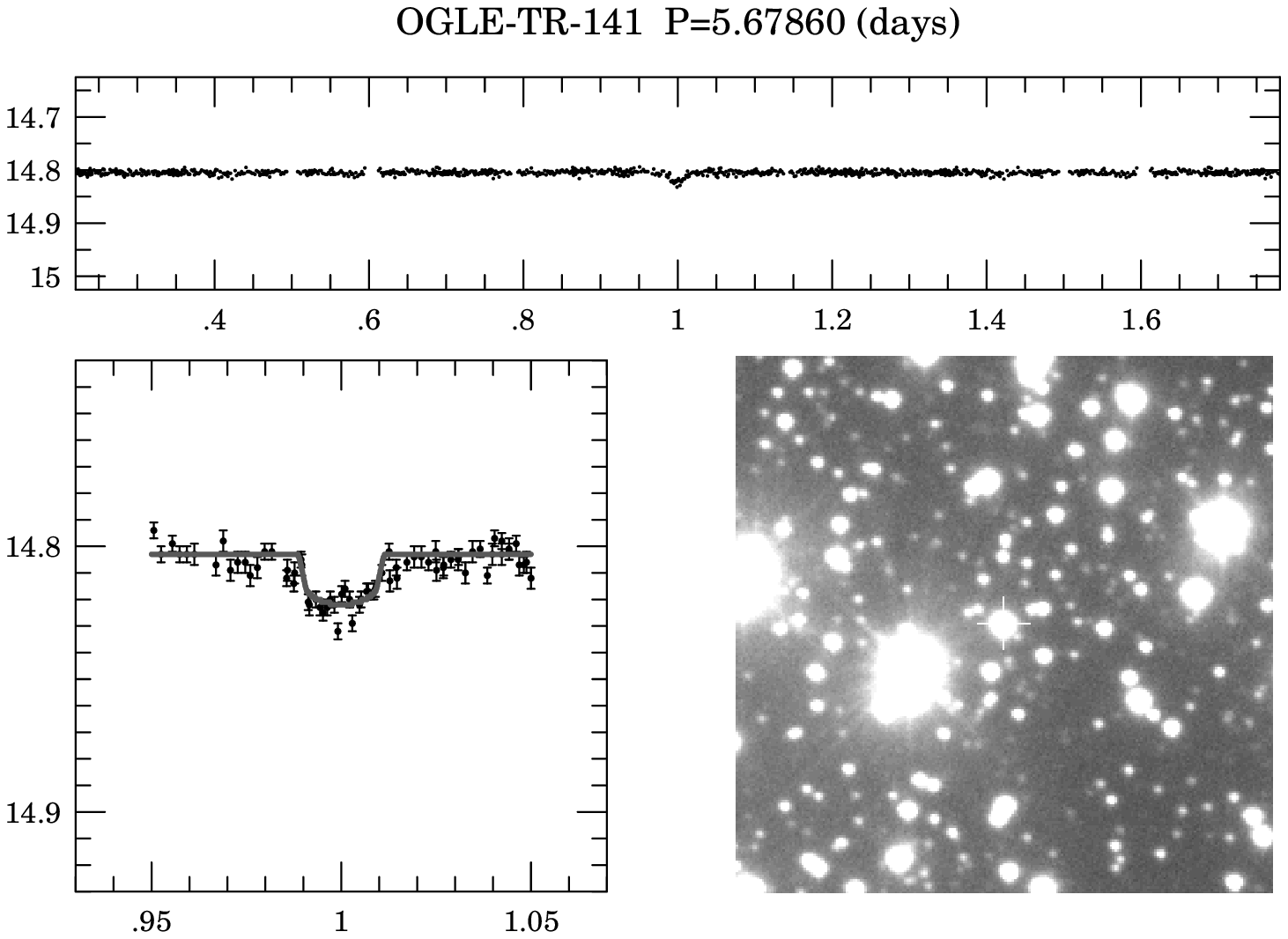}}
\end{figure} 
\clearpage
\begin{figure}[p] 
\vglue17mm
\centerline{\includegraphics[width=13cm, bb=100 140 540 460]{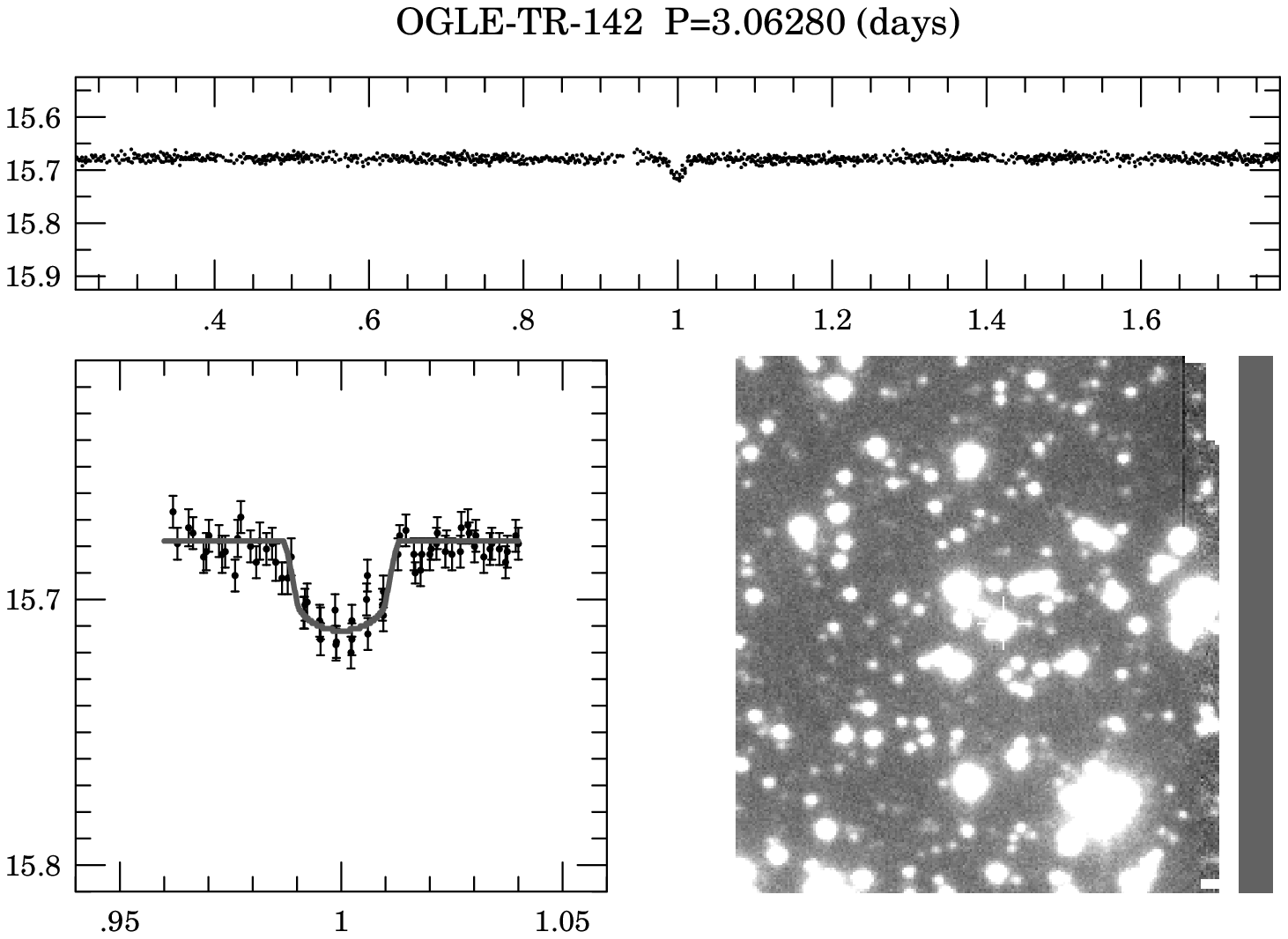}}
\vskip1.4cm
\centerline{\includegraphics[width=13cm, bb=100 140 540 460]{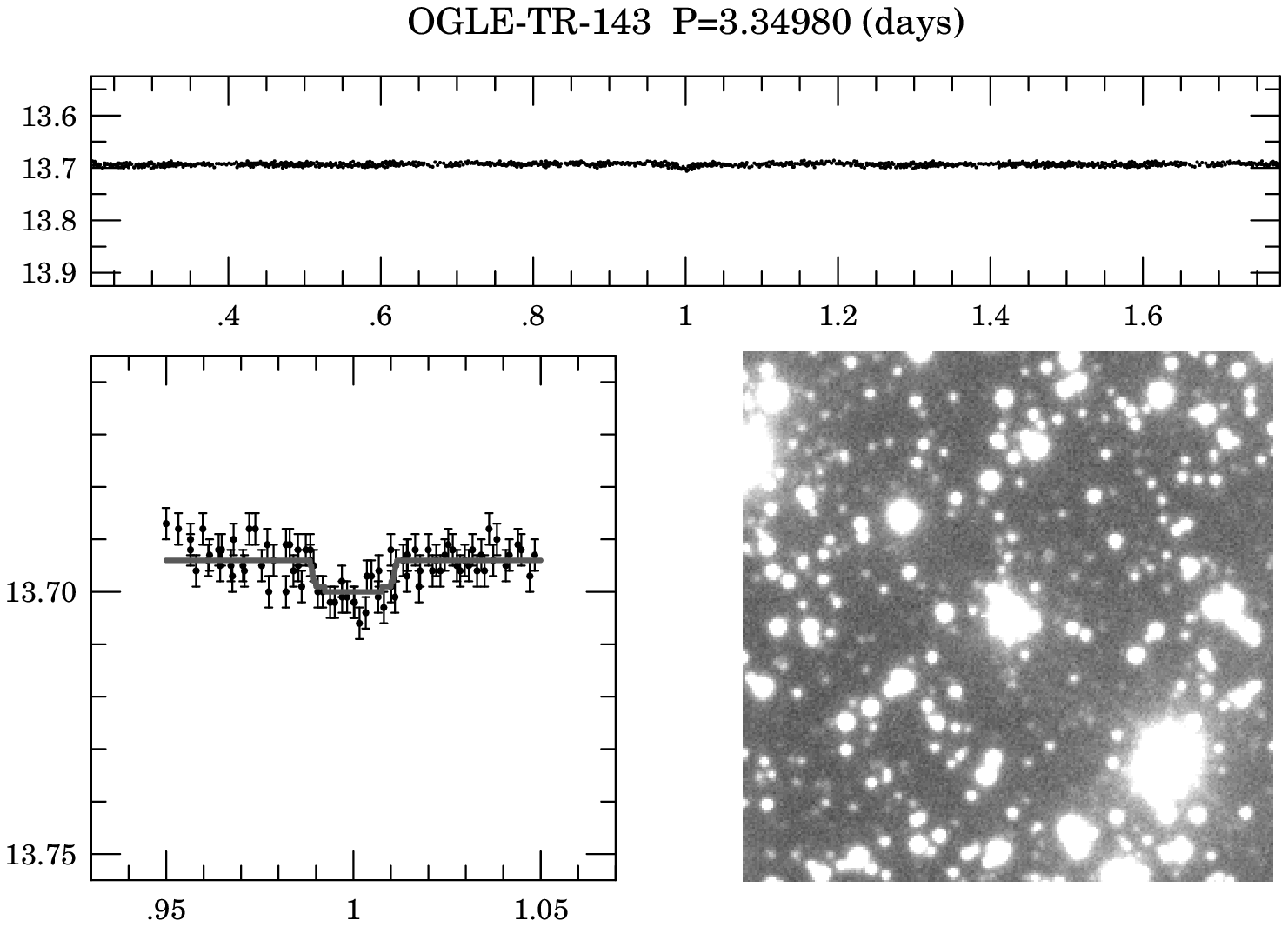}}
\end{figure} 
\end{document}